\documentclass{article}
\usepackage{interspeech2013, url, graphicx, amsmath, cite, balance}

\newif\ifniko
\nikotrue

\def\fig#1{Fig.~\ref{fig:#1}}
\def\sec#1{Section~\ref{sec:#1}}

\def\eq#1{(\ref{eq:#1})}

\let\underscore=\_
\def\_{\checkmath_\underscaore}
\def\checkmath#1#2{\ifmmode\def\next##1{#1{\rm##1}}\else\let\next=#2\fi\next}

\def\math#1{\relax\ifmmode#1\else$#1$\fi}

\def\N{{\cal N}}

\def\cdet{\math{C\_{det}}}

\def\cllr{\math{C\_{llr}}}

\def\mincllr{\math{C\_{llr}^{\rm min}}}

\def\model{\mathcal M}
\def\observ{\mathcal{O}}
\def\expect#1{\langle#1\rangle}
\def\Eset{\mathcal{E}}
\def\Dset{\mathcal{D}}
\def\cprimary{\math{C\_{primary}}}

\def\fudge{0.5}
\bigskipamount=\fudge\bigskipamount
\medskipamount=\fudge\medskipamount
\smallskipamount=\fudge\smallskipamount
\makeatletter
\renewcommand{\section}{\@startsection
  {section}% name
  {1}% level
  {}% indent
  {-\bigskipamount}%
  {0.5\bigskipamount}%
  {}}%

\renewcommand{\subsection}{\@startsection
  {subsection}%
  {2}%
  {}%
  {-\medskipamount}%
  {0.5\medskipamount}%
  {}}%

\renewcommand{\subsubsection}{\@startsection
  {subsubsection}%
  {3}%
  {}%
  {-\smallskipamount}%
  {0.5\smallskipamount}%
  {}}%
\makeatother

\renewenvironment{thebibliography}[1]{%
  \begin{oldthebibliography}{#1}%
    \setlength{\parskip}{0.3ex}%
    \setlength{\itemsep}{0.3ex}%
  }%
  {%
  \end{oldthebibliography}%
}

\title{The distribution of calibrated likelihood-ratios in speaker recognition}
\name{David A. van Leeuwen$^1$ and Niko Br\"ummer$^2$}
\address{$^1$Netherlands Forensic Institute, The Hague and Radboud University Nijmegen, The Netherlands\\
$^2$AGNITIO Research, Somerset West, South Africa}

\begin{document}

\maketitle
\begin{abstract}
\noindent
This paper studies properties of the score distributions of calibrated log-likelihood-ratios that are used in automatic speaker recognition.  We derive the essential condition for calibration that the log likelihood ratio of the log-likelihood-ratio is the log-likelihood-ratio.  We then investigate what the consequence of this condition is to the probability density functions (PDFs) of the log-likelihood-ratio score.  We show that if the PDF of the non-target distribution is Gaussian, then the PDF of the target distribution must be Gaussian as well.  The means and variances of these two PDFs are interrelated, and determined completely by the discrimination performance of the recognizer characterized by the equal error rate.  These relations allow for a new way of computing the offset and scaling parameters for linear calibration, and we derive closed-form expressions for these and show that for modern i-vector systems with PLDA scoring this leads to good calibration, comparable to traditional logistic regression, over a wide range of system performance. 
\end{abstract}

\ninept

\section{Introduction}

In recent years, calibration in automatic speaker recognition has received more attention~\cite{Brummer:2004,Brummer:2006,Brummer:2006a,Ramos:2006,Ramos:2007,ieee-stbu:2007,Jancik:2010,duration-calibration:2011,lr-lineup:2011,Miranti-icassp:2012,Doddington:2012}.  Intuitively, calibration is related to the ability to properly set a threshold in a speaker detection system so as to minimize the expected error~\cite{Doddington:2000}.  In speaker detection, the task is to decide whether or not two speech signals originate from the same speaker.  Because all speaker recognition systems internally work with some scalar \emph{score} that expresses speaker similarity, a score threshold can control the trade-off between the two types of errors that a system can make~\cite{Martin:1997, Appindep-eval:2007}. Indeed, in the series of NIST Speaker Recognition Evaluations (SRE) the primary evaluation measure has been sensitive to calibration.  Until SRE 2010, calibration was assessed in a single operating point, through a single decision cost function known as \cdet.  Also other technologies in speech technology or biometrics utilize calibration-sensitive evaluation measures, such as the cost functions $C\_{avg}$ in language recognition~\cite{Martin:2008} and the Half Total Error Rate in face recognition\cite{Wallace:2012}.

Since around 2004~\cite{Brummer:2004, Brummer:2006} the concept of calibration in speaker recognition has been generalized to a range of operating points by using proper scoring rules~\cite{DeGroot:1983} to evaluate probabilistic statements about whether a trial is a same-speaker (target) or different-speaker (non-target) trial.  A system that represents its score as a \emph{likelihood-ratio} can be well-calibrated over a wide range of operating points simultaneously.  This representation of the speaker recognition score has direct application in speaker detection, as the decision threshold follows directly from the cost function parameters~\cite{Appindep-eval:2007}, but also in evidence reporting in forensic speaker comparison cases~\cite{Ramos:2006,Gonzalez:2007}.  In the NIST SRE 2012, for the first time, hard decisions were no longer required, and instead the recognition score had to be submitted in the form of a likelihood-ratio.  The evaluation measure effectively sampled the decision cost function at two different parameters~\cite{nist-sre-evalplan:2012,plda-speakers:2013}.  

Since a calibrated likelihood-ratio is still just a score, all properties of normal scores apply to likelihood-ratios as well, and we can draw DET and ROC plots, determine EERs and inspect the score distributions.  The axis warping of the DET plot~\cite{Martin:1997} in combination with the observed more-or-less straight DET curves suggests that target and non-target score distributions could be accurately modelled with Gaussians.  These score distributions and the relation to the DET have been studied previously~\cite{Auckenthaler:2000,Navratil:2003} and are very instructive to the understanding of basic detection theory and the concepts of calibration~\cite{Appindep-eval:2007, Brummer-PhD:2010}.  In this paper we are interested in properties of the distributions of \emph{calibrated log-likelihood-ratios}. This may help situations were we carry out a calibration transformation on raw recognition scores, because it can tell us what the calibrated distributions should look like.     

The paper is organized as follows.  We define the very nature of a calibrated likelihood-ratio in \sec{definition}.  In \sec{gaussians} we investigate the properties of log-likelihood-ratio distributions when they are Gaussian, and we will then apply these in \sec{calibration} as a new method for calibration.  We then present experiments and conclusions. 

\section{Likelihood-ratio idempotence}
\label{sec:definition}

Here we carefully define the \emph{likelihood-ratio} (LR) and show that it has the interesting property: \emph{the LR of the LR is the LR}, which forms a definition of calibration.

The speaker recognition system has as input two speech segments, denoted $X$ and $Y$, which it processes in two steps. We represent the first step as $s=f(X,Y)$. To keep things general, $s$ may represent different kinds of output, e.g., a pair of acoustic feature vector sequences, a pair of i-vectors, or just a single, scalar recognition score.  The second step is to compute the likelihood-ratio~$r$ as a function of $s$, as:
\begin{align}
\label{eq:defLR}
r &= \frac{P(s\mid H_1,\model)}{P(s\mid H_2,\model)}
\end{align}
where $H_1$ is the (target) hypothesis that $X$ and $Y$ originate from the same speaker, $H_2$ the (non-target) hypothesis that they are from two different speakers, and $\model$ is a generative probabilistic model for $s$. In current practice, $s$ is always the recognition score, so that $\model$ merely models scalar scores---not i-vectors, acoustic feature sequences or speech signals. But our theory below is sufficiently general to remain applicable in future to more ambitious models, when $s$ might have a more complex form. We now assume there is given the \emph{hypothesis prior}, $\pi = P(H_1)$, which allows us to express the \emph{hypothesis posterior}, via Bayes' rule as:
\begin{align}
\label{eq:sufficiency}
P(H_1\mid s,\model,\pi) &= \frac{\pi r}{\pi r+ (1-\pi)}
\end{align}
This shows that $r$ is a \emph{sufficient statistic}: the posterior depends on $s$ only through $r$. This allows rewriting the posterior as:
\begin{align}
\label{eq:PostIsPost}
P(h\mid s,\model,\pi) = P(h\mid r,\model',\pi),\qquad h\in\{H_1,H_2\}
\end{align}
where we have introduced $\model'$ to denote $\model$, augmented by asserting~\eq{defLR}. Although $r$ contains all the relevant information that $\model$ can extract from $s$ to recognize the unknown hypothesis, it must be stressed that $r$ and $s$ do \emph{not} necessarily contain all the relevant information that could have been extracted from the original input $X,Y$ by some more elaborate model. Now we use the \emph{odds form} of Bayes' rule:
\begin{align}
\frac{P(H_1\mid \rho,M,\pi)}{P(H_2\mid \rho,M,\pi)} &= \frac{\pi}{1-\pi}\,\frac{P(\rho\mid H_1,M)}{P(\rho\mid H_2,M)}
\end{align}
where $\rho$ is a placeholder for $r$ or $s$ and $M$ for $\model$ or $\model'$. Combining this with~\eq{PostIsPost}, we find the desired relationship (the LR of the LR is the LR~\cite{Slooten:2012}):
\begin{align}
\label{eq:LRisLR}
r &= \frac{P(s\mid H_1,\model)}{P(s\mid H_2,\model)}
= \frac{P(r\mid H_1,\model')}{P(r\mid H_2,\model')}.
\end{align} 
If we define $x$ to be the \emph{log-likelihood-ratio} (LLR):
\begin{align}
\label{eq:defllr}
x &= \log r
\end{align}
we also find\footnote{To see this, note the log transformation is monotonic and the Jacobian of the transformation cancels in the ratio.} (the LLR of the LLR is the LLR):
\begin{align}
\label{eq:cal-llr-def}
x &= \log \frac{P(x\mid H_1,\model'')}{P(x\mid H_2,\model'')}
\end{align} 
where $\model''$ augments $\model'$ by addition of~\eq{defllr}.

\subsection{Implications}
Rewriting~\eq{LRisLR} as:
\begin{align}
P(r\mid H_1,\model') &= r P(r\mid H_2,\model')
\end{align} 
we see that if either of the two distributions is given, then the other distribution is completely determined---they cannot vary independently. Moreover, a further restriction is placed on these distributions: since the LHS must integrate to $1$, the \emph{expected value} of the non-target distribution (the integral of the RHS) must be: $\expect{r}=1$. Similarly, for targets: $\expect{\frac1r}=1$. By applying Jensen's inequality~\cite{Jensen:1906} we also find for targets: $\expect{x}\ge0$ and for non-targets: $\expect{x}\le0$.

\subsection{Good and bad calibration}

How does~\eq{LRisLR} function as a definition of calibration? Since it is an equality, won't all LRs calculated via~\eq{defLR} by some model $\model$, just automatically satisfy~\eq{LRisLR}? Yes they will, but only if $\model$ and $\model'$ are related as explained above. If we want to independently judge the goodness of the calibration of $r$, we do not condition the distributions for $r$ on the recognizer's model $\model$. Instead, we could empirically observe the target and non-target values of $r$ as calculated by the recognizer over an independent, supervised database of speaker detection trials. Letting $\observ$ denote the empirical observation, we could then say the model $\model$ is well calibrated if: 
\begin{align}
\label{eq:goodcal}
r &= \frac{P(s\mid H_1,\model)}{P(s\mid H_2,\model)}
\approx \frac{P(r\mid H_1,\observ)}{P(r\mid H_2,\observ)}
\end{align} 
Bad calibration is when the LRs given respectively by the recognizer's $\model$ and empirical observation $\observ$, do not agree in this way. This can and does happen, since $\observ$ is independent of any development data that was used to determine the form and parameters of $\model$.  

It should be noted that~\eq{goodcal} does not give a practical recipe to judge degree of goodness of calibration---it specifies neither how to assign $P(r\mid h,\observ)$, nor how to numerically evaluate the agreement between LHS and RHS. For practical solutions for calibration-sensitive objective functions, see for example~\cite{Brummer-Doddington:2013}.

\section{Gaussian distributed log-likelihood-ratios}
\label{sec:gaussians}

Inspired by the fact that DET curves in speaker recognition tend to be straight~\cite{Auckenthaler:2000}, we explore a Gaussian solution to the LLR distribution constraint~\eq{cal-llr-def}. Since target and non-target LLR distributions are so tightly coupled, it turns out that if the one is assumed to be Gaussian, then the other must also be. We shall use the shorthand: $e(x)=P(x\mid H_1,\model'')$ and $d(x)=P(x\mid H_2,\model'')$. Arbitrarily assuming a Gaussian distribution for non-targets (\textsl{different}-speaker trials):
\begin{equation}
  \label{eq:gaussian}
  d(x) = \N(x\mid \mu_d, \sigma_d) = \frac1{\sqrt{2\pi}\sigma_d} e^{-(x-\mu_d)^2/2\sigma_d^2}.
\end{equation}
We derive the functional form for targets\footnote{trials where the speakers are \textsl{equal}}, $e(x)$, when \eq{cal-llr-def} applies:
\begin{gather}
 e(x) = e^x d(x) = \frac1{\sqrt{2\pi}\sigma_d} e^{x-(x-\mu_d)^2/2\sigma_d^2}.\label{eq:e-expxd}
\end{gather}
We collect the terms in $x$ in the exponent, which itself can be written like
\begin{gather}
  \label{eq:6}
  -\frac{x^2 - 2\mu_dx + \mu_d^2}{2\sigma_d^2} + \frac{2\sigma_d^2x}{2\sigma_d^2} \\
 = -\frac{x^2 - 2(\mu_d+\sigma_d^2)x + \mu_d^2}{2\sigma_d^2} \\ 
 = -\frac{\bigl(x - (\mu_d+\sigma_d^2)\bigr)^2}{2\sigma_d^2} + \frac{2\mu_d\sigma_d^2 + \sigma_d^4}{2\sigma_d^2}
\end{gather}
The first term is in the familiar form of a Gaussian exponent, the second will result in a constant factor.  Gathering terms, and writing 
\begin{equation}
 \mu_e = \mu_d + \sigma_d^2,\label{eq:mu-and-sigma2-mu}
\end{equation}
the expression for the same-speaker comparison log-likeli\-hood-ratio scores becomes
\begin{align}
  \label{eq:8}
  e(x) &= \frac1{\sqrt{2\pi}\sigma_d} \, e^{\sigma_d^2/2 + \mu_d} \, e^{-(x-\mu_e^2)/2\sigma_d^2}\\
  &= e^{\sigma_d^2/2 + \mu_d} \,\N(x \mid \mu_e, \sigma_d).
\end{align}
We see that $e(x)$ is of Gaussian shape, with 
\begin{equation}
\label{eq:equal-sigma}
\sigma_e=\sigma_d \equiv \sigma.
\end{equation}
Since $e(x)$ must be a proper PDF, its integral over $x$ must be unity, from which follows that
\begin{gather}
  e^{\sigma^2/2 + \mu_d} \int_{-\infty}^\infty \N(x \mid \mu_e, \sigma)\,dx = 1\\
  -2\mu_d = \sigma^2. \label{eq:mu-d-and-sigma2}
\end{gather}
Finally, with \eq{mu-and-sigma2-mu} we find
\begin{equation}
 \mu_e = \mu_d + \sigma^2 = -\mu_d \equiv \mu,\label{eq:mu-and-mu}
\end{equation}
This shows that $d(x)$ and $e(x)$ are equal variance Gaussians with means symmetric around zero at $\pm\mu$, and where the variance and mean are related~\eq{mu-d-and-sigma2}
\begin{align}
  \label{eq:mu-and-sigma2}
  \sigma^2=2\mu.
\end{align}

\ifniko\else
In \fig{2gaussians} the position of the two distributions is drawn.  

\begin{figure}
  \centering
  \includegraphics[width=\hsize]{2gaussians.pdf}
  \vglue-\baselineskip
  \caption{The solution for Gaussian PDFs $d(x)$ (left) and $e(x)$ (right).}
  \label{fig:2gaussians}
\end{figure}
\fi

\subsection{Equal Error Rate and $d'$}

Using the symmetry of the solution, it is clear that the threshold for the equal error rate is at $x=0$.  Using the expression for the miss probability, the equal error rate $E_=$ is 
\begin{align}
  E_= &= \int_{-\infty}^0 \N(x\mid \mu, \sigma)\\
  &= \int_{-\infty}^{-\mu/\sigma} \N(x\mid 0, 1) \equiv \Phi(-\mu/\sigma),   \label{eq:eer-int}
\end{align}
where $\Phi(x)$ is the cumulative normal distribution. 

It is sometimes useful to recognize the parameter $d'$ from detection theory, which is the difference in means expressed in terms of the standard deviation, here $d'=2\mu/\sigma$.  With \eq{eer-int} the relation becomes
\begin{align}
  E_= &= \Phi(-\frac12d').\label{eq:eer}\\
  d' = \sigma &= -2 \Phi^{-1}(E_=),\label{eq:dprime}
\end{align}
introducing $\Phi^{-1}(y)$, the inverse of the cumulative normal distribution.   The importance of the relations above is that $\mu$ and~$\sigma$ are determined by the discrimination performance measured by~$E_=$, using \eq{mu-and-sigma2} and~\eq{dprime} 
\begin{align}
  \label{eq:mu-from-eer}
  \mu = \frac{\sigma^2}2 = 2[\Phi^{-1}(E_=)]^2.
\end{align}

\section{A new calibration method}
\label{sec:calibration}

In practice, automatic speaker recognition systems do not deliver scores that can directly be interpreted as a log-likelihood-ratio, even though they are computed as such, for instance in the good old UBM-GMM scoring~\cite{Reynolds:2000} or the latest i-vector PLDA scoring~\cite{Prince:2007}.  A practical solution to this is to convert raw scores~$s(X,Y)$ to calibrated log-likelihood-ratios by some transformation function $x(s)$, usually constrained to be monotonic increasing.  There are many ways of doing this. The FoCal~\cite{FoCal} and BOSARIS~\cite{bosaris:2010} toolkits use logistic regression to discriminatively train linear calibration transformations. Other possibilities include isotonic regression (PAV~\cite{bosaris:2010}) and line-up calibration~\cite{lr-lineup:2011} that uses the rank in a line-up of foil speakers. In FoCal or BOSARIS, the score-to-LLR function is affine:
\begin{align}
\label{eq:affine}
  x(s) = as+b
\end{align}
and the parameters $a$ and~$b$ are found by optimizing cross-entropy, a calibration-sensitive objective function defined on a supervised set of speaker recognition trials.  

Here we contrast the popular discriminative logistic regression solution to a new generative, constrained maximum-likelihood (ML) solution. Our constraints follow from assuming (i) Gaussian LLR distributions, and (ii) an affine score-to-LLR transform~\eq{affine}. This implies that (i) the LLR distributions are constrained as derived in \sec{gaussians}, and (ii) the score distributions are also Gaussians, with equal variances. With no LLR distribution constraints, we would have had 6 free parameters: 2 means, 2 variances and 2 calibration parameters. But we have imposed 3 constraints, equal variances~\eq{equal-sigma}, symmetric means~\eq{mu-and-mu} and~\eq{mu-and-sigma2}. We find the remaining 3 free parameters by maximizing the following weighted likelihood:
\begin{align*}
\frac{\alpha}{N_e}\sum_{i\in\Eset} \log\N(s_i\mid m_e,v) 
+\frac{1-\alpha}{N_d}\sum_{i\in\Dset} \log\N(s_i\mid m_d,v) 
\end{align*}
where $\Eset$ and $\Dset$ index $N_e$ target, and $N_d$ non-target scores, weighted by $\alpha$ and $1-\alpha$, respectively. The score distribution parameters that need to be optimized are the means $m_e,m_d$ and common variance $v$. Setting derivatives to $0$, we find the maximum likelihood at the sample means:
\begin{align}
m_e = \frac{1}{N_e}\sum_{i\in\Eset}s_i, \qquad
m_d = \frac{1}{N_d}\sum_{i\in\Dset}s_i
\end{align}
and at a weighted combination of sample variances:
\begin{align}
\label{eq:weighted-variance}
v = 
\frac{\alpha}{N_e}\sum_{i\in\Eset}(s_i-m_e)^2
+\frac{1-\alpha}{N_d}\sum_{i\in\Dset}(s_i-m_d)^2
\end{align}
By~\eq{affine}, the LLR distribution parameters become $\sigma^2=a^2v$, $\mu_e=am_e+b$ and $\mu_d=am_d+b$. Finally, applying the constraints $\sigma^2=\mu_e-\mu_d$ and $\mu_e=-\mu_d$, we can solve for the calibration parameters:
\begin{align}
\label{eq:CMLG-cal}
a = \frac{m_e-m_d}{v},\qquad b = -a\frac{m_e+m_d}{2} 
\end{align}
We call this recipe \emph{constrained, maximum-likelihood, Gaussian} (CMLG) calibration.  An advantage of CMLG is that it has a closed form, in contrast to the iterative optimization required by logistic regression.

\subsection{Experiment}

In order to test CMLG we apply it to a number of recognition trials sets.  We use a set of trials crafted for duration-dependence experiments~\cite{duration-calibration:2011} from the NIST SRE 2008 and 2010 trial sets, the telephone-telephone ``extended'' trial lists.  We constructed short duration segments of 5, 10, 20, and 40 seconds from both train and test segments by simply selecting the first frames after speech activity detection.  All durations, including the full conversation side, were tested in all combinations, leading to 25 different trial lists.  The NIST SRE10 `det-5' performance over these lists ranges from $E_==2.9$--$26$\,\%.  The recognition system is a standard i-vector based system with PLDA scoring described elsewhere~\cite{plda-speakers:2013}. 

We contrast CMLG (with $\alpha=\frac12$) to the traditional logistic regression method. The calibrations are trained on NIST SRE 2008 data (427\,375 trials) and applied to SRE 2010 trials for evaluation (10\,007\,900 trials), all gender mixed.  We evaluate the 25 different trial list combinations using \cllr, a cost function that is sensitive to calibration over the whole DET curve~\cite{Brummer:2006}.  We used R's \texttt{glm} routine for logistic regression. 

\begin{figure}
%% at isti:~/werk/calibration
%% x <- cal.all.durdata()
%% with(x, plot(ideal.cllr, logreg.cllr, pch=16, cex=1.5, xlab="constrained maximum likelihood Gaussian (CMLG)", ylab="logistic regression", main="Correlation in Cllr for different calibration methods", panel.first=grid()))
  \centering
  \vglue-\baselineskip
  \includegraphics[width=\hsize]{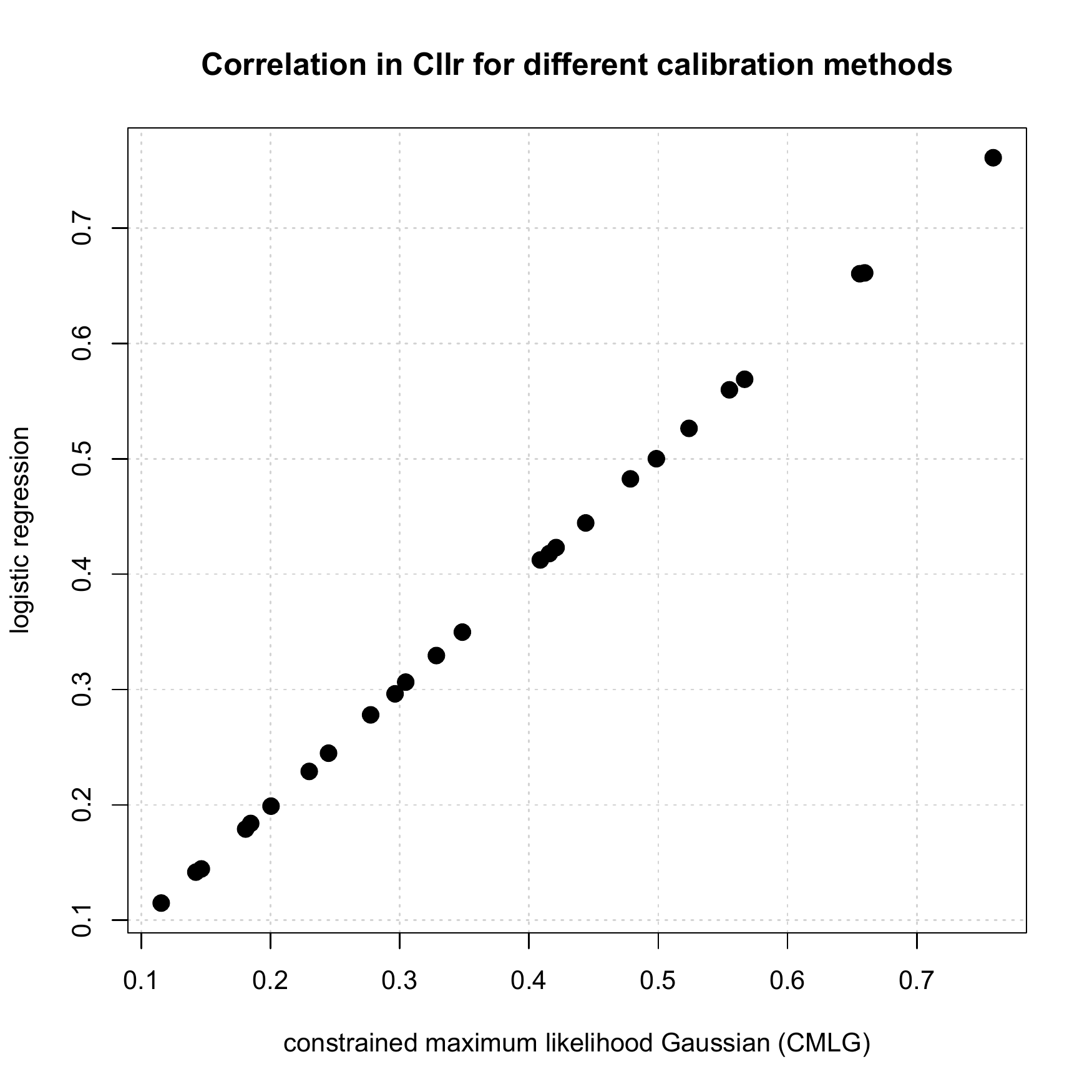}
  \vglue-1.5\baselineskip
  \caption{\protect\cllr\ values of the 25 trial lists for the CMLG method (horizontal) versus logistic regression (vertical).  }
 \vglue-\baselineskip
  \label{fig:cllr}  
\end{figure}

The results are shown in \fig{cllr}, where we have plotted the \cllr\ obtained using CMLG calibration versus \cllr\ obtained using logistic regression.  The values are highly correlated.  For CMLG, the average \cllr\ over all 25 conditions is 0.375, for logistic regression it is 0.376.  These can be called good, as the mean \mincllr\ is 0.370.  

\ifniko
We have also used the NIST SRE12 scores from the ABC-team to study the effect of $\alpha$ in \eq{weighted-variance} to another calibration sensitive measure~\cprimary, cf.~\fig{niko-prior}, for details we refer to~\cite{Brummer-Doddington:2013}.  The figure shows that with CMLG good calibration results can be obtained for a different system with different data and a different performance measure, if the correct~$\alpha$ is chosen.  
\begin{figure}
%% n <- read.table("niko-prior-scan.txt", header=T)
%% with(n, plot(log(alpha)-log(1-alpha), logreg, type="b", xlab=expression(log(alpha) - log(1-alpha)), ylab=expression(C[primary]), lwd=2, col="red", ylim=c(0.27,0.60), main="Comparison of calibration methods", pch=1, panel.first=grid()))
%% with(n, lines(log(alpha)-log(1-alpha), CMLG, type="b", lwd=2, col="blue", pch=2))
 \centering
  \vglue-\baselineskip
  \includegraphics[width=\hsize]{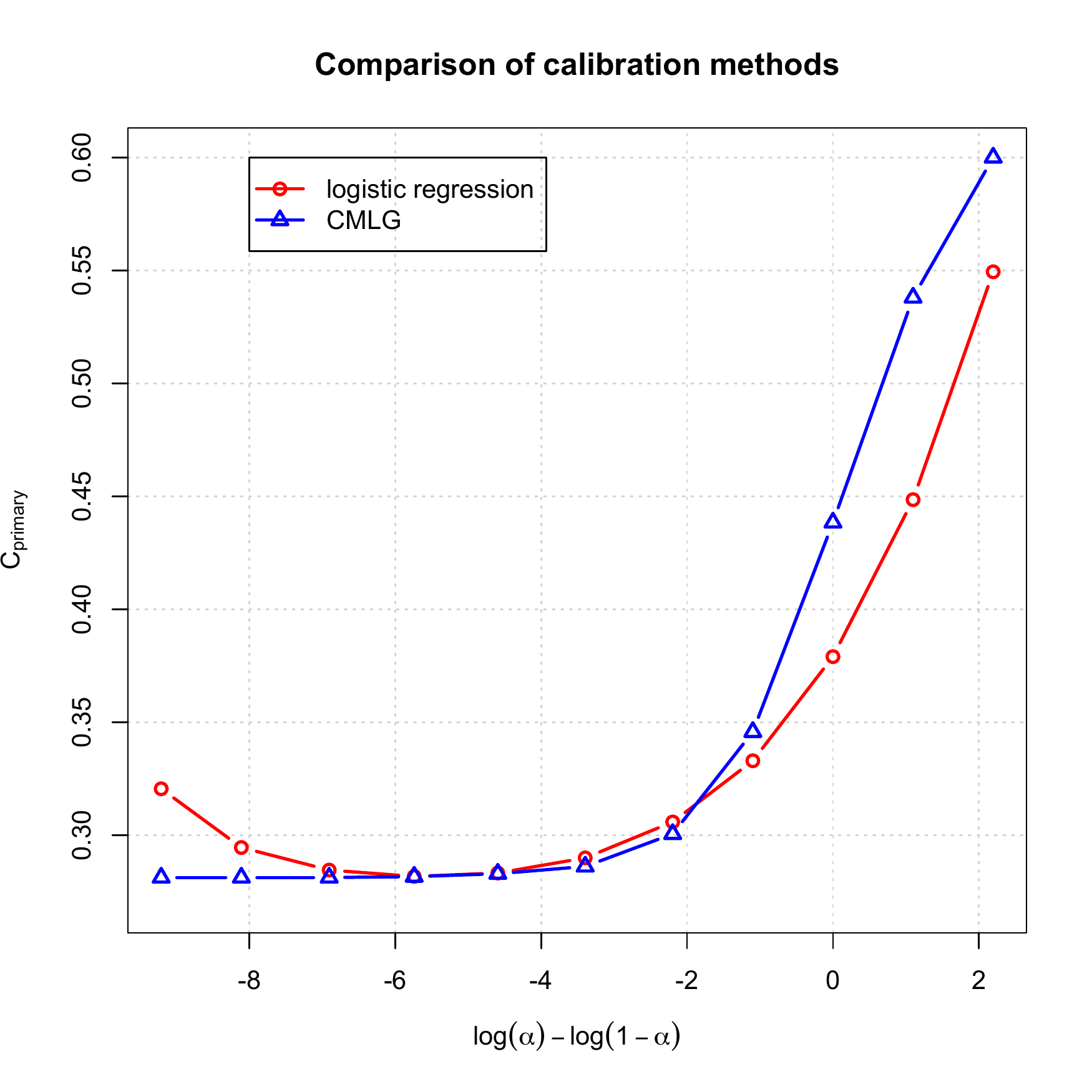}
  \vglue-1.5\baselineskip  
  \caption{\protect\cprimary\ for logistic regression and CMLG calibration methods for ABC's SRE12 submission, as a function of prior~$\alpha$ used in the objective / ML optimization. }
  \label{fig:niko-prior}
  \vglue-\baselineskip
\end{figure}
\fi

\section{Discussion and Conclusions}

We have shown in this paper, that if the different-speaker calibrated log-likelihood-ratio scores from a speaker recognition system follow a Gaussian distribution, then the distribution of the same-speaker scores must also be Gaussian after calibration, with the same variance but opposite mean. Because monotonically  increasing score-to-likelihood-ratio functions do not change the DET plot, such equal-variance distributions in the calibrated score domain imply $45^\circ$ DET-plots in the raw score domain as well---which is neither observed with real data\footnote{We have measured the slope of the DET in the conventional error region 0.1--50\,\% for the data in the experiment.  The mean slope over the 25 conditions is $-0.99$ with a standard deviation of $0.06$, so in fact this data appears to honour the equal variance condition quite well.} nor desired for applications operating in the low false alarm region.  The logical conclusion then is that real scores, if they are well-calibrated, will not be Gaussian.  However, we see that our PLDA system can be calibrated quite well under the Gaussian assumptions, and indeed we have noticed that i-vector PLDA systems tend to have score distributions that appear more Gaussian than earlier technologies, such as i-vector LDA cosine distance scoring, support vector machines or the UBM-GMM likelihood ratio scoring.  

The Gaussian solution to the LLR equation~\eq{cal-llr-def} is one where both distributions are shaped by the same mathematical function.  In signal detection theory, where the distribution represents noise, this seems almost mandatory, but in speaker recognition this is not an obvious assumption.  We have experimented with other distributions, e.g., in the likelihood-ratio domain~\eq{LRisLR} a pair of Gamma distributions is a solution to the calibration condition, and these are asymmetric in the log-likelihood-ratio domain.  However, such distributions seem to be not at all representative of real score distributions.  Also, an arbitrary linear combination of Gaussians with different means and corresponding variances is a solution to~\eq{cal-llr-def} which allows some freedom in fitting a shape of score distribution.  In principle, there is no need for real score distributions to follow any mathematical description, but we have observed that many researchers like to use some form of idealized shape of the score distributions to understand the data~\cite{Auckenthaler:2000,Ramos:2006}.  When calibration methods are designed, condition~\eq{cal-llr-def} should therefore be taken into account. 

The relations derived in \sec{gaussians} open up more possibilities for relations between the various evaluation measures.  For instance, we can compute \cllr\ by numerical integration as 
\begin{equation}
  \label{eq:12}
  \frac1{\log 2}\int_{-\infty}^\infty \N(x\mid\nobreak \mu, \sigma) \log(1+e^{-x})\,dx 
\end{equation}
and this relates \cllr\ to $E_=$ via~\eq{dprime} and~\eq{mu-from-eer} for Gaussian score distributions.  E.g., for our set of 25 trial lists this expression differs from \mincllr\ only 0.006 in root mean squared difference, or about 2\,\%. 
\ifniko
Instead of for calibration, the relations can also be used for fusion of systems.  For pre-calibrated systems this leads to solutions that transparently depend on the correlation between the scores. 
\else
Further, the expectation properties of the LR in \sec{definition} can be utilized for yet another form of calibration by finding parameters to \eq{affine} that minimize the empirical error in the expectation expressions, with some regularization.  
\fi

The fact that we can obtain the linear calibration parameters under the Gaussian assumption is an interesting side-effect of this study.  The calibration parameters can be expressed in closed-form, and do not explicitly consider cross entropy or \cllr\ as an optimization objective.  For score distributions that do not resemble a Gaussian, this calibration method is likely to fail---we therefore do not recommend CMLG calibration as a general technique.  Still, we are quite pleased that the experiments support the mostly theoretical results of this paper.   

\section{Acknowledgments}
The research leading to these results has received funding from the European Community's Seventh Frame work Programme (FP7/2007--2013) under grant agreement no.~238803.

\balance
\ninept
\let\_=\underscore

\bibliographystyle{IEEEtran}
\bibliography{david-bibdesk}

\begin{thebibliography}{10}
\providecommand{\url}[1]{#1}
\csname url@samestyle\endcsname
\providecommand{\newblock}{\relax}
\providecommand{\bibinfo}[2]{#2}
\providecommand{\BIBentrySTDinterwordspacing}{\spaceskip=0pt\relax}
\providecommand{\BIBentryALTinterwordstretchfactor}{4}
\providecommand{\BIBentryALTinterwordspacing}{\spaceskip=\fontdimen2\font plus
\BIBentryALTinterwordstretchfactor\fontdimen3\font minus
  \fontdimen4\font\relax}
\providecommand{\BIBforeignlanguage}[2]{{%
\expandafter\ifx\csname l@#1\endcsname\relax
\typeout{** WARNING: IEEEtran.bst: No hyphenation pattern has been}%
\typeout{** loaded for the language `#1'. Using the pattern for}%
\typeout{** the default language instead.}%
\else
\language=\csname l@#1\endcsname
\fi
#2}}
\providecommand{\BIBdecl}{\relax}
\BIBdecl

\bibitem{Brummer:2004}
N.~Br\"ummer, ``Application-independent evaluation of speaker detection,'' in
  \emph{Proc.\ Odyssey 2004 Speaker and Language recognition workshop}.\hskip
  1em plus 0.5em minus 0.4em\relax ISCA, June 2004, pp. 33--40.

\bibitem{Brummer:2006}
N.~Br\"ummer and J.~du~Preez, ``Application-independent evaluation of speaker
  detection,'' \emph{Computer Speech and Language}, vol.~20, pp. 230--275,
  2006.

\bibitem{Brummer:2006a}
N.~Br\"ummer and D.~A. van Leeuwen, ``On calibration of language recognition
  scores,'' in \emph{Proc.\ Odyssey 2006 Speaker and Language recognition
  workshop}, San Juan, June 2006.

\bibitem{Ramos:2006}
D.~Ramos-Castro, J.~Gonz\'alez-Rodr\'{\i}guez, and J.~Ortega-Garcia,
  ``Likelihood ratio calibration in a transparent and testable forensic speaker
  recognition framework,'' in \emph{Proc.\ Odyssey 2006 Speaker and Language
  Recognition Workshop}, 2006.

\bibitem{Ramos:2007}
D.~Ramos, ``Forensic evaluation of the evidence using automatic speaker
  recognition systems,'' Ph.D. dissertation, Universidad Autonoma de Madrid,
  November 2007.

\bibitem{ieee-stbu:2007}
N.~Br\"ummer, L.~Burget, J.~\v{C}ernock\'y, O.~Glembek, F.~Grezl,
  M.~Karafi\'at, P.~Mat\v{e}jka, D.~A. van Leeuwen, P.~Schwarz, and
  A.~Strassheim, ``Fusion of heterogeneous speaker recognition systems in the
  {STBU} submission for the {NIST} speaker recognition evaluation 2006,''
  \emph{IEEE Transactions on Speech, Audio and Language Processing}, vol.~15,
  no.~7, pp. 2072--2084, 2007.

\bibitem{Jancik:2010}
Z.~Jancik, O.~Plchot, N.~Brummer, L.~Burget, O.~Glembek, V.~Hubeika,
  M.~Karafi{\'a}t, P.~Matejka, T.~Mikolov, A.~Strasheim \emph{et~al.}, ``Data
  selection and calibration issues in automatic language
  recognition--investigation with {BUT-AGNITIO} {NIST} {LRE} 2009 system,'' in
  \emph{Proc.\ Speaker and Language Odyssey}, 2010.

\bibitem{duration-calibration:2011}
M.~I. Mandasari, M.~McLaren, and D.~A. van Leeuwen, ``Evaluation of i-vector
  speaker recognition systems for forensic application,'' in \emph{Proc.\
  Insterspeech}.\hskip 1em plus 0.5em minus 0.4em\relax Firenze: ISCA, August
  2011.

\bibitem{lr-lineup:2011}
D.~A. van Leeuwen and N.~Br\"ummer, ``A speaker line-up for the likelihood
  ratio,'' in \emph{Proc.\ Interspeech}.\hskip 1em plus 0.5em minus 0.4em\relax
  Firenze: ISCA, August 2011.

\bibitem{Miranti-icassp:2012}
M.~I. Mandasari, M.~McLaren, and D.~A. van Leeuwen, ``The effect of noise on
  modern automatic speaker recognition systems,'' in \emph{Proc.\
  ICASSP}.\hskip 1em plus 0.5em minus 0.4em\relax Kyoto: IEEE, March 2012.

\bibitem{Doddington:2012}
G.~R. Doddington, ``The role of score calibration in speaker recognition,'' in
  \emph{Proc.\ Interspeech}, 2012.

\bibitem{Doddington:2000}
G.~R. Doddington, M.~A. Przybocki, A.~F. Martin, and D.~A. Reynolds, ``The
  {NIST} speaker recognition evaluation---{O}verview, methodology, systems,
  results, perspective,'' \emph{Speech Communication}, vol.~31, pp. 225--254,
  2000.

\bibitem{Martin:1997}
A.~Martin, G.~Doddington, T.~Kamm, M.~Ordowski, and M.~Przybocki, ``The {DET}
  curve in assessment of detection task performance,'' in \emph{Proc.\
  Eurospeech 1997}, Rhodes, Greece, 1997, pp. 1895--1898.

\bibitem{Appindep-eval:2007}
D.~A. van Leeuwen and N.~Br\"ummer, ``An introduction to
  application-independent evaluation of speaker recognition systems,'' in
  \emph{Speaker Classification}, ser. {Lecture Notes in Computer Science /
  Artificial Intelligence}, C.~M\"uller, Ed.\hskip 1em plus 0.5em minus
  0.4em\relax Springer, 2007, vol. 4343.

\bibitem{Martin:2008}
A.~F. Martin and A.~N. Le, ``{NIST} 2007 language recognition evaluation,'' in
  \emph{Proc.\ Speaker and Language Odyssey}.\hskip 1em plus 0.5em minus
  0.4em\relax Stellenbosch, South Afrika: IEEE, 2008.

\bibitem{Wallace:2012}
R.~Wallace, M.~McLaren, C.~McCool, and S.~Marcel, ``Cross-pollination of
  normalization techniques from speaker to face authentication using gaussian
  mixture models,'' \emph{Information Forensics and Security, IEEE Transactions
  on}, vol.~7, no.~2, pp. 553--562, 2012.

\bibitem{DeGroot:1983}
M.~DeGroot and S.~Fienberg, ``The comparison and evaluation of forecasters,''
  \emph{The Statistician}, pp. 12--22, 1983.

\bibitem{Gonzalez:2007}
J.~Gonzalez-Rodriguez, P.~Rose, D.~Ramos, D.~T. Toledano, and J.~Ortega-Garcia,
  ``Emulating {DNA}: Rigorous quantification of evidential weight in
  transparent and testable forensic speaker recognition,'' \emph{IEEE
  Transactions on Audio, Speech and Language Processing}, vol.~15, no.~7, pp.
  2104--2115, September 2007.

\bibitem{nist-sre-evalplan:2012}
\BIBentryALTinterwordspacing
C.~S. Greenberg, ``The {NIST} year 2012 speaker recognition evaluation plan,''
  2012. [Online]. Available:
  \url{http://www.nist.gov/itl/iad/mig/upload/NIST_SRE12_evalplan-v17-r1.pdf}
\BIBentrySTDinterwordspacing

\bibitem{plda-speakers:2013}
D.~A. van Leeuwen and R.~Saeidi, ``Knowing the non-target speakers: the effect
  of the i-vector population for {PLDA} training in speaker recognition,'' in
  \emph{Proc ICASSP}.\hskip 1em plus 0.5em minus 0.4em\relax Vancouver: IEEE,
  2013.

\bibitem{Auckenthaler:2000}
R.~Auckenthaler, M.~Carey, and H.~Lloyd-Thomas, ``Score normalization for
  text-independent speaker verification systems,'' \emph{Digital Signal
  Processing}, vol.~10, pp. 42--54, 2000.

\bibitem{Navratil:2003}
J.~{Navr\' atil} and G.~N. Ramsawamy, ``The awe and mistery of t-norm,'' in
  \emph{Proc.~Eurospeech}, 2003, pp. 2009--2012.

\bibitem{Brummer-PhD:2010}
N.~Br\"ummer, ``Measuring, refining and calibrating speaker and language
  information extracted from speech,'' Ph.D. dissertation, Stellenbosch
  University, 2010.

\bibitem{Slooten:2012}
K.~Slooten and R.~Meester, ``Forensic identification: Database likelihood
  ratios and familial {DNA} searching,'' \emph{arXiv:1201.4261 [stat.AP]},
  2012.

\bibitem{Jensen:1906}
J.~L. W.~V. Jensen, ``Sur les fonctions convexes et les in\'egalit\'es entre
  les valeurs moyennes,'' \emph{Acta Mathematica}, vol.~30, no.~1, pp.
  175--193, 1906.

\bibitem{Brummer-Doddington:2013}
N.~Br\"ummer and G.~Doddington, ``Likelihood-ratio calibration using
  prior-weighted proper scoring rules,'' in \emph{Proc.\ Interspeech}.\hskip
  1em plus 0.5em minus 0.4em\relax ISCA, 2013.

\bibitem{Reynolds:2000}
D.~A. Reynolds, T.~F. Quatieri, and R.~B. Dunn, ``Speaker verification using
  adapted gaussian mixture models,'' \emph{Digital Signal Processing}, vol.~10,
  pp. 19--41, 2000.

\bibitem{Prince:2007}
S.~J.~D. Prince and J.~H. Elder, ``Probabilistic linear discriminant analysis
  for inferences about identity,'' in \emph{IEEE International Conference on
  Computer Vision (ICCV)}.\hskip 1em plus 0.5em minus 0.4em\relax IEEE, 2007,
  pp. 1--8.

\bibitem{FoCal}
N.~Br\"ummer, \emph{Fo{C}al-II: Toolkit for calibration of multi-class
  recognition scores}, August 2006, software available at
  \url{http://www.dsp.sun.ac.za/~nbrummer/focal/index.htm}.

\bibitem{bosaris:2010}
E.~de~Villiers and N.~Br\"ummer, \emph{The Bosaris Toolkit}, BOSARIS, 2010,
  software available at \url{https://sites.google.com/site/bosaristoolkit/}.

\end{thebibliography}

\end{document}